\documentclass[baaa]{baaa}

\usepackage[pdftex]{hyperref}
\usepackage{subfigure}
\usepackage{natbib}
\usepackage{helvet,soul}
\usepackage[font=small]{caption}

\contriblanguage{1}

\contribtype{1}

\thematicarea{3}

\title{Origin of magnetism in early-type stars}

\titlerunning{Origin of magnetism in early-type stars}

\author{
J.P. Hidalgo\inst{1}, 
P.J. Käpylä\inst{2,3}, 
C.A. Ortiz-Rodríguez\inst{1}, 
F.H. Navarrete\inst{4}, 
B. Toro \inst{1} 
\& 
D.R.G. Schleicher\inst{1}
}

\authorrunning{Hidalgo et al.}

\contact{jhidalgo2018@udec.cl}

\institute{
Departamento de Astronomía, Universidad de Concepción, Chile
\and
Institut für Astrophysik und Geophysik, Georg-August-Universität Göttingen, Alemania 
\and
Nordita, KTH Royal Institute of Technology and Stockholm University, Suecia
\and
Hamburger Sternwarte, Universität Hamburg, Alemania}

\resumen{De acuerdo a nuestro entendimiento de la evolución estelar, las estrellas de tipo temprano poseen envolturas radiativas y núcleos convectivos debido a un fuerte gradiente de temperatura producido por el cíclo CNO. Algunas de estas estrellas (principalmente, las subclases Ap y Bp) tienen fuertes campos magnéticos, lo suficiente para ser observados por el efecto Zeeman. Aquí, presentamos simulaciones magnetohidrodinámicas en 3D de una estrella tipo A con $2~\mathrm{M}_{\odot}$ utilizando el modelo star-in-a-box. Nuestra meta es explorar si la estrella modelada es capaz de mantener un campo magnético tan fuerte como los observados, a través de un dínamo en su núcleo convectivo, o manteniendo una configuración estable de un campo fósil proveniente de una etapa evolucionaria temprana, usando diferentes velocidades de rotación. Creamos dos modelos, uno parcialmente radiativo y otro totalmente radiativo, que están determinados por el valor de la conductividad térmica. Nuestro modelo es capaz de explorar ambos escenarios, con dínamos relevantes impulsados por convección.}

\abstract{According to our understanding of stellar evolution, early-type stars have radiative envelopes and convective cores due to a steep temperature gradient produced by the CNO cycle. Some of these stars (mainly, the subclasses Ap and Bp) have strong magnetic fields, enough to be directly observed using the Zeeman effect. Here, we present 3D magnetohydrodynamic simulations of an $2 ~\mathrm{M}_{\odot}$ A-type star using the star-in-a-box model. Our goal is to explore if the modeled star is able to maintain a magnetic field as strong as the observed ones, via a dynamo driven by its convective core, or via maintaining a stable fossil field configuration coming from its early evolutionary stages, using different rotation rates. We created two models, a partially radiative and a fully radiative one, which are determined by the value of the heat conductivity. Our model is able to explore both scenarios, including convection-driven dynamos.
}

\keywords{ Stars: magnetic fields --- stars: massive --- magnetohydrodynamics (MHD) --- dynamo }

\begin{document}

\maketitle

\section{Introduction}\label{S_intro}

Magnetic fields are ubiquitous in the universe, and there is a general consensus that they are amplified and maintained via astrophysical dynamos. In stars, these processes typically require rotation and convection, and therefore are most likely to occur inside convection zones (see \citealt{Brandenburg-2005}). Main-sequence stars with masses above $\sim 1.5~\mathrm{M}_{\odot}$ have stably stratified radiative envelopes and convective cores due to a steep temperature gradient produced by the CNO cycle. The least massive spectral type that fulfils these characteristics are A-type stars, which in general tend to be fast rotators \citep{Royer-2007} and have very weak magnetic fields of the order of a few Gauss. Interestingly, there is a clear bimodality here (see \citealt{Auriere-2007}), the peculiar-subclass Ap stars have slow rotation rates and magnetic fields between $300 ~\mathrm{G}$ and $30~\mathrm{kG}$, with the highest one so far reaching $\sim 34 ~\mathrm{kG}$ \citep{Babcock-1960}. The origin of these magnetic fields remains uncertain, but there are some theories: one includes a very strong core dynamo. This in principle could create a large scale magnetic field in the surface if it is strong enough, but also requires an efficient transport mechanism \citep{Moss-1989}. \cite{Augustson-2016} performed 3D simulations of a $10 ~\mathrm{M}_{\odot}$ B-type star, modeling the inner 64\% of its radius excluding the innermost values of the core to avoid a coordinate singularity. They found core dynamos able to produce strong magnetic fields, with peak strengths exceeding a megagauss. Another theory is that the magnetic field of these stars is a fossil field, a remnant from an earlier evolutionary stage that has survived in a stable configuration. Simulations made by \cite{Braithwaite-2006} of a $2~\mathrm{M}_\odot$ A-type star, have found stable axisymmetric magnetic field configurations starting with random field initial conditions. Also, non-axisymmetric configurations were found starting from turbulent initial conditions \citep{Braithwaite-2008}. 

The aim of our project is to explore both scenarios mentioned above. We explain our methods, initial conditions and the model in Section \ref{S_model}, the preliminary results in Section \ref{results}, and finally, a brief conclusion followed by the planned future work in Section \ref{future}.

 
\begin{figure}[h!]
\centering
\includegraphics[width=\columnwidth]{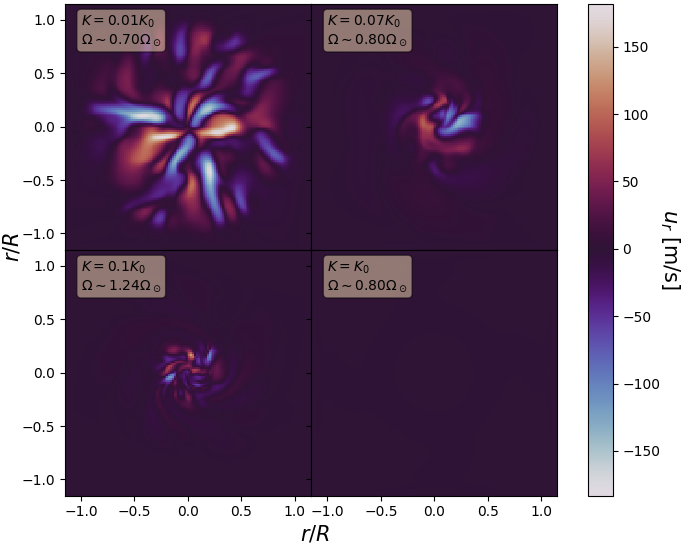}
\caption{Snapshots of star-in-a-box simulations, showing the equatorial plane. The values of the heat conductivity $K$ (where $K_0$ is the value for a fully radiative configuration) and the rotation rate $\Omega$ (in $\Omega_\odot$) are shown in each plot. The colorbar represents the radial component of the flow velocity, where regions with $u_r \neq 0$ are convection zones.}
\label{Figure1}
\end{figure}

\section{The Model}\label{S_model}
We use a star-in-a-box set-up based on the model presented by \cite{kapyla-2021} with a star of radius $R$ inside a Cartesian cube of side $l=2.2R$ where all coordinates $(x,y,z)$ range from $-l/2$ to $l/2$. The set of magnetohydrodynamics (MHD) equations is the following:
\begin{align}
    \frac{\partial \mathbf{A}}{\partial t} &= \mathbf{u} \times \mathbf{B} - \eta \mu_0 \mathbf{J}, \label{mhd_1} \\
    \frac{D \ln \rho}{D t} &= - \boldsymbol{\nabla} \cdot \mathbf{u}, \label{mhd_2} \\
    \frac{D \mathbf{u}}{D t} &= - \boldsymbol{\nabla} \Phi - \frac{1}{\rho} \left( \boldsymbol{\nabla} p - \boldsymbol{\nabla} \cdot 2 \nu \rho \mathbf{S} + \mathbf{J} \times \mathbf{B} \right) \nonumber \\
    &\hspace{4.3cm} - 2 \mathbf{\Omega}\times \mathbf{u} + \mathbf{f}_d, \label{mhd_3} \\
    T \frac{Ds}{Dt} &= - \frac{1}{\rho} \left[ \boldsymbol{\nabla}\cdot (\mathbf{F}_\text{rad} + \mathbf{F}_\text{SGS})  + \mathcal{H} - \mathcal{C} + \mu_0 \eta \mathbf{J}^2 \right] \nonumber \\
    &\hspace{4.3cm} + 2 \nu \mathbf{S}^2, \label{mhd_4}
\end{align}
where $\mathbf{A}$ is the magnetic vector potential, $\mathbf{u}$ is the flow velocity, $\mathbf{B} = \boldsymbol{\nabla} \times \mathbf{A}$ is the magnetic field, $\eta$ is the magnetic diffusivity, $\mu_0$ is the magnetic permeability of vacuum, $\mathbf{J} = \boldsymbol{\nabla} \times \mathbf{B}/\mu_0$ is the current density given by Ampère's law, $D/Dt = \partial/\partial t + \mathbf{u} \cdot \boldsymbol{\nabla}$ is the advective (or material) derivative, $\rho$ is the mass density, $\Phi$ is the gravitational potential corresponding to the isentropic hydrostatic state of an A0 star, $p$ is the pressure, $\mathbf{S}$ is the traceless rate-of-strain tensor, $T$ is the temperature, $\mathbf{\Omega} = (0,0,\Omega_0)$ is the rotation rate along the $z$ axis, $\mathbf{f}_\mathrm{d}$ describes damping of flows exterior to the star. Radiation inside the star is approximated as a diffusion process. Therefore, the radiative flux is given by:
\begin{align}
    \mathbf{F}_\mathrm{rad} = - K \boldsymbol{\nabla} T, \label{radiative-flux}
\end{align}
where $K$ is the radiative heat conductivity, a quantity that is assumed to have a constant profile and establishes the size of the radiative zone (see Figure \ref{Figure1}). In addition, it is convenient to introduce a subgrid-scale (SGS) entropy diffusion that does not contribute to the net energy transport, but damps fluctuations near the grid scale. This is given by the SGS entropy flux $\mathbf{F}_\mathrm{SGS} = -\chi_\mathrm{SGS} \rho \boldsymbol{\nabla}s'$, where $s'$ is the fluctuating entropy.

Finally, $\mathcal{H}$ and $\mathcal{C}$ describe additional heating and cooling (respectively), and we adopted similar expressions as \cite{dobler-2006} and \cite{kapyla-2021}. \\

The simulations were run on a grid of $128^3$ using the {\sc Pencil Code}, a highly modular high-order finite-difference code for compressible non-ideal MHD \citep{pencilcode}. The stellar parameters used for a $2~\mathrm{M}_{\odot}$ A0-type star are $R_* = 2~\mathrm{R}_\odot$, $L_* = 23~\mathrm{L}_\odot$, $\rho_* \approx 5.6 \cdot 10^{4} ~\mathrm{kg}\,\mathrm{m}^{-3}$ for the radius, the luminosity and the central mass density respectively, which were obtained using the open-source stellar evolution code {\sc MESA} (see \citealt{MESA-2011}). For the relation to reality and the treatment of the units, we followed the description in Appendix A of \cite{kapyla-units}.

\begin{table}[t!]
\centering
\caption{Summary of all runs. $\Delta r$ denotes the radial extent of the convective core (where $R$ is the stellar radius), $K_0$ is the value for a fully radiative configuration, $\nu$ and $\eta$ in [$\mathrm{m}^2\,\mathrm{s}^{-1}$], $\Omega$ in [$\Omega_\odot$], $\langle u_\mathrm{rms} \rangle$ in [$\mathrm{m}\,\mathrm{s^{-1}}$], and $B_\mathrm{max}$ (the maximum value of $B_\mathrm{rms}$) in [$\mathrm{kG}$].}
\begin{tabular}{lcccccc}
\hline\hline\noalign{\smallskip}
\!\!Run & \!\!\!\!$K/K_0$ & \!\!\!\!$\nu~[10^{9}]$& \!\!\!\!$\eta~[10^{9}]$ &\!\!\!\!$\Omega$&\!\!\!\!$\langle u_\mathrm{rms}\rangle$ & $B_\mathrm{max}$
\\
\hline\noalign{\smallskip}
\!\!Sim1  &  $0.01$ & $2$ & $1$ & 0.14 & 127 & 65\\
\!\! \!\!$\Delta r \approx 1R$& $0.01$ & $2$ & $1$ & 0.70 & 82 & 50\\
\hline\noalign{\smallskip}
\!\!Sim2 & $0.04$ & $1.2$ & $1$ & 0.10 & 118 & 49\\
\!\!$\Delta r \approx 1R$ & $0.04$ & $1.2$ & $1$ & 0.20 & 95 & 65\\
\hline\noalign{\smallskip}
\!\!Sim3 & $0.07$ & $0.2$ & $0.18$ & 0.80 & 284 & 25\\
\!\!$\Delta r \approx 0.3R$ & $0.07$ & $0.2$ & $0.18$ & 1.58 & 229 & 27\\
\hline\noalign{\smallskip}
\!\!Sim4 & $0.1$ & $0.12$ & $0.18$ & 1.24 & 264 & 24\\
\!\!$\Delta r \approx 0.2R$ & $0.1$ & $0.12$ & $0.18$ & 2.48 & 177 & 24\\
\!\! & $0.1$ & $0.12$ & $0.18$ & 6.20 & 162 & 22\\
\hline
\end{tabular}
\label{tabla1}
\end{table}
\section{Preliminary results}\label{results}
\begin{figure*}[h!]
\centering
\includegraphics[scale=0.44]{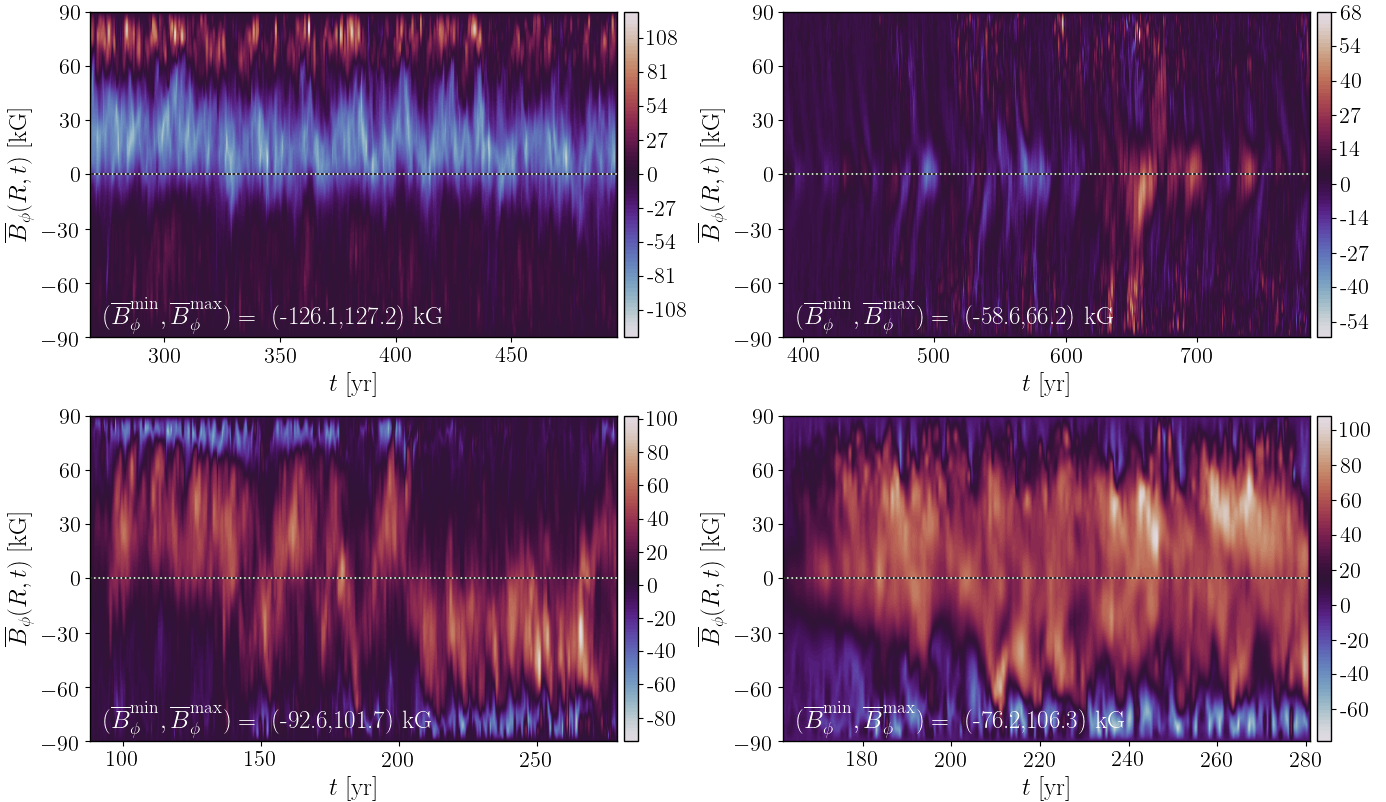}
\caption{Azimuthally averaged magnetic field [kG] vs time [year]. The \emph{upper panels} correspond to Sim1, with $\Omega = 0.14~\Omega_\odot$ (\emph{left}) and $\Omega = 0.70~\Omega_\odot$ (\emph{right}). The \emph{lower panels} are the runs from Sim2, $\Omega = 0.10~\Omega_\odot$ (\emph{left}) and $\Omega = 0.20~\Omega_\odot$ (\emph{right}).}
\label{Figure2}
\end{figure*}

The simulations are listed in Table \ref{tabla1}, divided into 4 main groups with different values for the diffusivities $\nu$, $\eta$, and the radiative heat conductivity $K$ which determines the depth of the convective zone $\Delta r$. The averages for the root-mean-square flow velocity $\langle u_\mathrm{rms} \rangle$ are estimated considering motions inside the convection zone. The rotation rates were chosen in order to have the Coriolis number
\begin{align}
    \mathrm{Co} = \frac{2\Omega_0}{u_\mathrm{rms} k_R}, \label{eq-Co}
\end{align}
equal to $\mathrm{Co} \approx 1$, $\mathrm{Co} \approx 2$, and $\mathrm{Co} \approx 10$ (only in Sim4), where $k_R = 2\pi/\Delta r$ corresponds to the scale of the largest convective eddies.

The simulations Sim1 and Sim2 are fully convective ($\Delta r \approx 1R$), which is not realistic for a main-sequence A-type star; however, these scenarios are useful as a way to test the model and can be representative of pre-main sequence evolution. Figure \ref{Figure2} shows the time evolution of the azimuthaly averaged magnetic field $\Bar{B}_\phi$ on the stellar surface. It is possible to find very strong magnetic fields, even though diffusivity values are quite high. We found quasi-steady solutions like in the upper-left and lower-right panels, and more interestingly, a polarity change in the upper-right panel.

\begin{figure}[h!]
\centering
\includegraphics[width=\columnwidth]{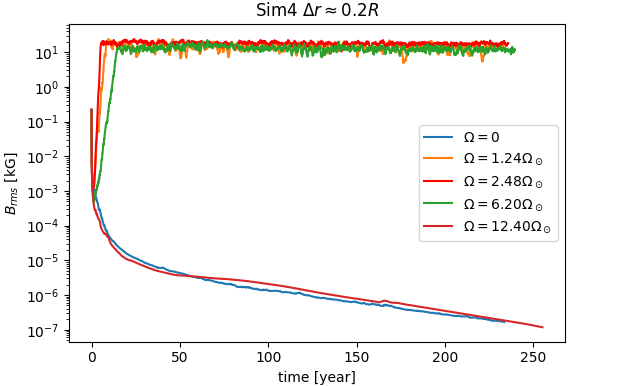}
\caption{Root-mean-square magnetic field $B_\mathrm{rms}$ [kG] vs time [year], of the runs from Sim4.}
\label{Figure3}
\end{figure}

The partially convective runs also generate very strong dynamos inside their cores. Root-mean-square magnetic fields from Sim4 can be seen in Figure \ref{Figure3}. Here, we also included two runs (non-rotating and very rapid rotation) that were not dynamos, and therefore were not included in Table \ref{tabla1}. The run with $\Omega = 6.20~\mathrm{\Omega}_\odot$ has the highest amplitude $\Bar{B}_\phi$ of the group at $0.2R$, with $(\Bar{B}_\phi^{\mathrm{min}}, \Bar{B}_\phi^{\mathrm{max}}) = (-197.0,216.7)~\mathrm{kG}$. Sim3 behaves similarly to Sim4, where we obtain peak $B_\mathrm{rms}$ values around $20-30~\mathrm{kG}$, and the run with $\Omega = 1.58~\mathrm{\Omega}_\odot$ has $(\Bar{B}_\phi^{\mathrm{min}}, \Bar{B}_\phi^{\mathrm{max}}) = (-146.9,134.8)~\mathrm{kG}$, which corresponds to the highest field amplitude in the group at $0.3R$.

\section{Conclusions and future work}\label{future}

We explored different scenarios for an A-type star, with convective cores of $100 \%$, $30 \%$ and $20 \%$ of stellar radius. Our model is able to generate magnetic fields in all of them, and the $20 \%$ case, which is the most realistic, has also the highest value of the azimuthally averaged magnetic field. The current results are promising, but Sim3 and Sim4 need to be analyzed more carefully.



\begin{acknowledgement}
We gratefully acknowledge support by the ANID BASAL projects ACE210002 and FB210003, as well as via Fondecyt Regular (project code 1201280).
\end{acknowledgement}


\bibliographystyle{baaa}
\small
\bibliography{777_v1.bib}

\end{document}